\documentstyle[11pt,paspconf,epsf]{article}

\begin{document}

\def\ltsima{$\buildrel<\over\sim$}
\def\lsim{\lower.5ex\hbox{\ltsima}}
\def\gtsima{$\buildrel>\over\sim$}
\def\gsim{\lower.5ex\hbox{\gtsima}}


 \title{Modeling  age  and   metallicity   in  globular   clusters:  a
 comparison of theoretical giant branch isochrones}

\author{T.  Lejeune and R.  Buser}
\affil{Astron.\  Institut der Universit\"at Basel, CH-4102 Binningen,
 Switzerland}

\begin{abstract}
In view of the important contributions of red giants to the integrated
light  in   evolutionary   synthesis   calculations  of   old  stellar
populations  in clusters and   galaxies,  we provide a comparison   of
theoretical isochrones computed   from  the Padova,  Geneva,  and Yale
evolutionary tracks, respectively.  Using the  most recent version  of
the color-calibrated library of theoretical stellar spectra by Lejeune
et al.  (1999), the isochrones  are converted in  a uniform manner  to
the observational  color-magnitude  diagrams, ($M(T_1)$,  $C-T_1$) and
($M(I)$, $V-I$), and compared  with observed giant branches of typical
globular  clusters spanning a  large  range of metallicities.  We find
that the  three different isochrones of  the red  giant branch provide
significantly different slopes  and curvatures, where in general,  the
differences tend  to  be  larger for   metallicities  decreasing below
$[M/H] \sim -1$.  Throughout  the full metallicity range ($-2$ {\lsim}
$[M/H]$ {\lsim} $0$), the Yale isochrones are the only ones which show
very  good  matches  of both    the observed  giant  branches and  the
associated metallicity scale derived from the color  of the RGB, while
the   Padova   isochrones   exhibit  significant  discrepancies    for
low-metallicities.  The effects of these differences on the integrated
model colors of single stellar  populations, in particular for age and
metallicity determinations, are briefly addressed.
\end{abstract}


\section{Introduction}
Evolutionary population synthesis  (EPS) is now a traditional approach
to study distant galaxies from their integrated light.  This technique
has  its   foundation in   stellar evolution  theory   which,  with an
appropriate library of stellar spectra, allows to predict the spectral
evolution  of  a  given  stellar population.   In   this  context, the
globular   clusters as single      stellar populations (SSP)   of  old
metal-deficient   stars   put  stringent   constraints on evolutionary
synthesis  models, which are particularly  useful in investigating the
effects of age and metallicity  in the integrated  spectro-photometric
properties.  In view of    the important contributions of red    giant
branch  (RGB)  stars to  the integrated   light (up  to  40--60 \%) in
evolutionary  synthesis   calculations of old   stellar populations in
clusters   and  galaxies, it is  therefore  important  to  compare the
predictions of  different sets  of  theoretical isochrones  frequently
used in EPS  studies, such  as those  computed  from  the Padova  (P),
Geneva (G), and Yale (Y) evolutionary tracks.

\section{Theoretical color--absolute magnitude diagrams}

In this work, we used the Padova (P) isochrones ($Z=0.0001$ to $0.10$)
as calculated from the {\em Isochrone Synthesis} program of Bruzual \&
Charlot  (see  Bruzual  et   al.   1997),   the  Yale   (Y) isochrones
($Z=0.0002$ to $0.10$) from Demarque et al.  (1996) and the Geneva (G)
isochrones  ($Z=0.001$ to  $0.10$)  calculated by Schaerer  \& Lejeune
(1998).  All the isochrones  have been converted to  the observational
color--magnitude (c-m)  diagrams,   ($M(T_1)$,  $C-T_1$) and  ($M(I)$,
$V-I$), in a uniform  manner, {\it i.e.},  by employing the {\em same}
stellar library  to transform  the theoretical quantities ($M_{\mathrm
bol}$, $T_{\rm eff}$)  into absolute magnitudes  and colors.  For this
purpose, we used the most recent version of  the Basel stellar library
({\it  B}a{\it S}e{\it L}) of   Lejeune et al.   (1999) which provides
{\em color-calibrated}  theoretical flux distributions for the largest
possible range of fundamental stellar  parameters, $T_{\rm eff}$ (2000
K to 50,000 K), $\log g$  (-1.0 to 5.5) , and  $[Fe/H]$ (-5.0 to +1.0)
(for details, see Lejeune et al.  1997, 1998, 1999, and Westera et al.
1999 -- these proceedings).

\begin{figure}
\plotone{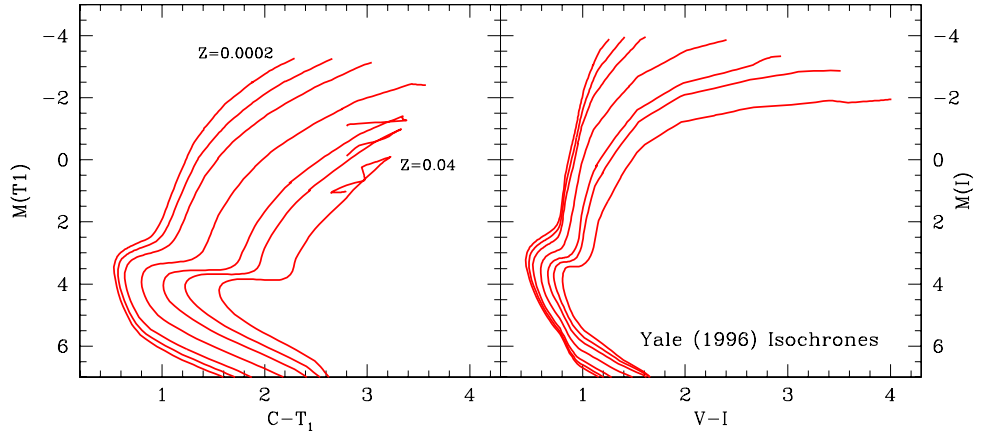}
 \caption{A comparison  of  the  theoretical isochrones for  different
 abundances ($Z=0.0002$ to $0.04$)  in the ($M(T_1)$, $C-T_1$)  and in
 the  ($M(I)$,  $V-I$)  c-m     diagrams   (left and right     panels,
 respectively) .}
\end{figure}

In Figure 1, we provide a comparison of the c-m diagrams computed from
the Y-isochrones in the Washington and in the Johnson-Cousins systems,
($M(T_1)$,  $C-T_1$)  and   ($M(I)$, $V-I$),   respectively.  Both the
position  and the  curvature  of the  RGB  are  very sensitive  to the
metallicity, as already noticed in the empirical studies by Da Costa
\& Armandroff  (1990, hereafter DCA90)  for $V-I$,  and by  Geisler \&
Sarajedini (1999, [GS99])  for $C-T_1$.  The  larger separation of the
isochrones in $C-T_1$ confirms the superior metallicity sensitivity of
the Washington system, in  particular at low-metallicities.  Note also
that the  observed  {\em  near-constancy of the   absolute magnitude},
$M(I)$ or $M(T_1)$,  of the RGB-tip,   is confirmed by theory  for $Z$
{\lsim} 0.04  ($[M/H]$    {\lsim}  $-0.5$)  to   within  a  dispersion
$\sigma(M)$ {\lsim}  1 mag -- supporting  the use of this  property in
Galactic and extragalactic distance determinations.


\section{Comparisons with observed giant branches}

\begin{figure}
\plotone{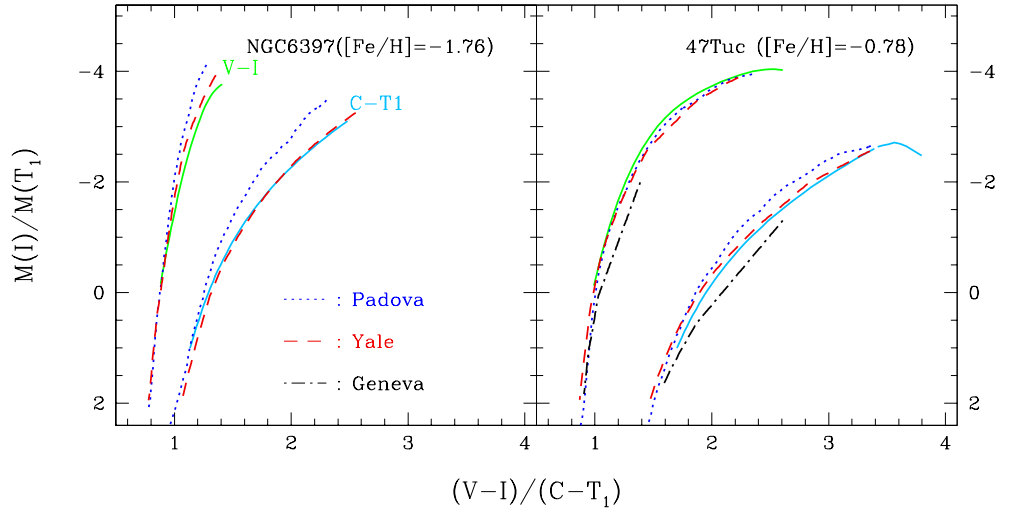}
 \caption{Theoretical  RGBs  compared to standard-observed RGBs (solid
 lines) for  two clusters given by DCA90  in $M(I)$  vs.  $V-I$ and by
 GS99 in $M(T_1)$ vs.  $C-T_1$.  The isochrones have been interpolated
 graphically  (linearly in [Fe/H])  at  the metallicity of the cluster
 (as given by Caretta \& Gratton, 1997),  assuming a typical age of 14
 Gyr.   Enhancement from the  $\alpha$--elements,  when known for  the
 cluster, has been taken  into account following the  prescriptions of
 Salaris et al.  (1993).  Note the large discrepancy in the upper part
 of the    RGB between  the   P-models and  the   observations for the
 metal-poor globular cluster NGC 6357.}
\end{figure}

In Figure 2, theoretical RGBs from the P-, the Y- and the G-isochrones
at  a typical  age of  14  Gyr  are  compared with  the observed  {\em
standard} giant branches of the globular  clusters NGC 6397 and 47 Tuc
defined by  DCA90   in $V-I$  and by  GS99  in   $C-T_1$.   Systematic
differences exist between the  theoretical isochrones and the observed
giant branches,  and  between the   models themselves.  While  the  P-
isochrones reproduce very well the RGB  of 47 Tuc, large discrepancies
are found  for NGC 6357 ($[Fe/H]=-1.76$),  both in  $V-I$ and $C-T_1$.
More generally, our tests with other clusters  (Lejeune \& Buser 1999)
indicate  that the P-models  provide   reliable theoretical RGBs   for
metal-rich and intermediate-metallicity  globular clusters, while they
become  systematically too blue for  $[M/H]$  {\lsim} $-1.4$.  This is
particularly true  in Washington  photometry  with color  differences,
$\Delta  (C-T_1)$, at the   RGB-tip increasing up  to  $0.4$  mag with
decreasing   $[M/H]$.   At the  opposite,  the   Geneva  models appear
slightly too red when compared to the different cluster branches.  Our
comparisons show  in   particular   that, over   the  whole range   of
metallicities $(\sim -2 <[M/H] < 0)$ and within an age range between 8
and 17 Gyr, the best agreement is  provided by the Y-isochrones, which
reproduce very well the  position  and the  curvature of  the observed
standard giant branches of DCA90  and GS99,  along with the  empirical
metallicity scales derived  from the  color of the    RGB at a   fixed
absolute magnitude (Lejeune \& Buser 1999).

\section{Conclusions}

In  view of their use   in  old stellar  population  studies, we  have
compared  the predictions of Padova, the  Yale and the  Geneva sets of
theoretical isochrones   with recent observations of  globular cluster
giant branches. We find in particular that, at a given typical old age
of 14 Gyr, the 3 different isochrones of the RGB provide significantly
different slopes   and  curvatures, with   differences  increasing for
metallicities decreasing below $[M/H]$ {\lsim} $-1$.  The Y-isochrones
are the only  ones which show excellent  agreement throughout the full
range of metallicities   with observed globular  cluster  giant-branch
templates  in both photometric  systems.   The P-isochrones agree well
with observations only   at  the higher metallicities,  while  for the
lower metallicities  they become generally  too blue, by up  to $-0.4$
mag  at   the  tip  of  the   RGB.   Finally,   the  G-isochrones  are
systematically redder than   the observations. As all the  theoretical
isochrones  have been converted in an  uniform manner by employing the
same stellar library, such discrepancies  must likely be attributed to
differences in the physics  employed in the different  calculations of
the stellar evolutionary tracks.

The influence of  such differences on  the integrated colors predicted
by  stellar  population   models,  which can lead   in   particular to
systematic  uncertainties   in  age   ($\sim$ 2--3  Gyr)   and/or   in
metallicity ($\sim$ 0.2--0.3 dex), will be  discussed in a forthcoming
paper (Lejeune \& Buser 1999).

\acknowledgments

T. L.   gratefully  acknowledges  financial  support  from the   Swiss
National Science Foundation (grant  20-53660.98  to Prof. Buser),  and
from  the  ``Soci\'et\'e  Suisse  d'Astronomie   et  d'Astrophysique''
(SSAA).

\end{document}